\documentclass[prl,twocolumn,superscriptaddress,amsmath,amssymb,floatfix,noshowpacs]{revtex4}
\usepackage{graphicx,hyperref}
\usepackage{color}
\usepackage{soul}
\usepackage{amsmath}
\usepackage{tabularx}
\graphicspath{{fig/}}

\newcommand{\hide}[1]{}

\begin{document}
\title{\textbf{Cavity quantum optomechanics with an atom-array membrane}}
%\date{\today}
\author{Ephraim Shahmoon}
\affiliation{Department of Chemical \& Biological Physics, Weizmann Institute of Science, Rehovot 761001, Israel}
\affiliation{Department of Physics, Harvard University, Cambridge, Massachusetts 02138, USA}
\author{Dominik S.~Wild}
\affiliation{Department of Physics, Harvard University, Cambridge, Massachusetts 02138, USA}
\author{Mikhail D.~Lukin}
\affiliation{Department of Physics, Harvard University, Cambridge, Massachusetts 02138, USA}
\author{Susanne F.~Yelin}
\affiliation{Department of Physics, Harvard University, Cambridge, Massachusetts 02138, USA}
\affiliation{Department of Physics, University of Connecticut, Storrs, Connecticut 06269, USA}
\date{\today}

\begin{abstract}
We consider a quantum optomechanical scheme wherein an ordered two-dimensional array of laser-trapped atoms is used as a movable membrane. The extremely light mass of the atoms yields very strong optomechanical coupling, while their spatial order largely eliminates scattering losses. We show that this combination opens the way for quantum optomechanical nonlinearities, well within the ultimate single-photon strong-coupling regime. As an example, we analyze the possibility to observe optomechanically induced quantum effects such as photon blockade and time-delayed non-classical correlations. We discuss novel opportunities opened by the optomechanical backaction on the internal states of the array atoms.
\end{abstract}
 \maketitle

The field of quantum optomechanics deals with nonlinear phenomena formed by the backaction of radiation pressure on light. It is typically studied in a cavity setting, where light that circulates in the cavity pushes a reflector, e.g. an internal membrane, whose position determines the cavity resonance frequency and hence the phase of the light (Fig. 1a) \cite{AKM,MEY,DOR,HAR}. This optomechanical nonlinearity plays an important role in various quantum technologies \cite{CAV,LIGO,LEH}, and has lead to great advancements such as ground-state cooling of massive objects \cite{MAR,WIL,SCH,CHAN,TEU} and optomechanical quantum squeezing \cite{HAM,SK3,REG,SAF,SCH2,KIP,SIL1,SIL2}.

\begin{figure}[t]
  \begin{center}
    \includegraphics[width=\columnwidth]{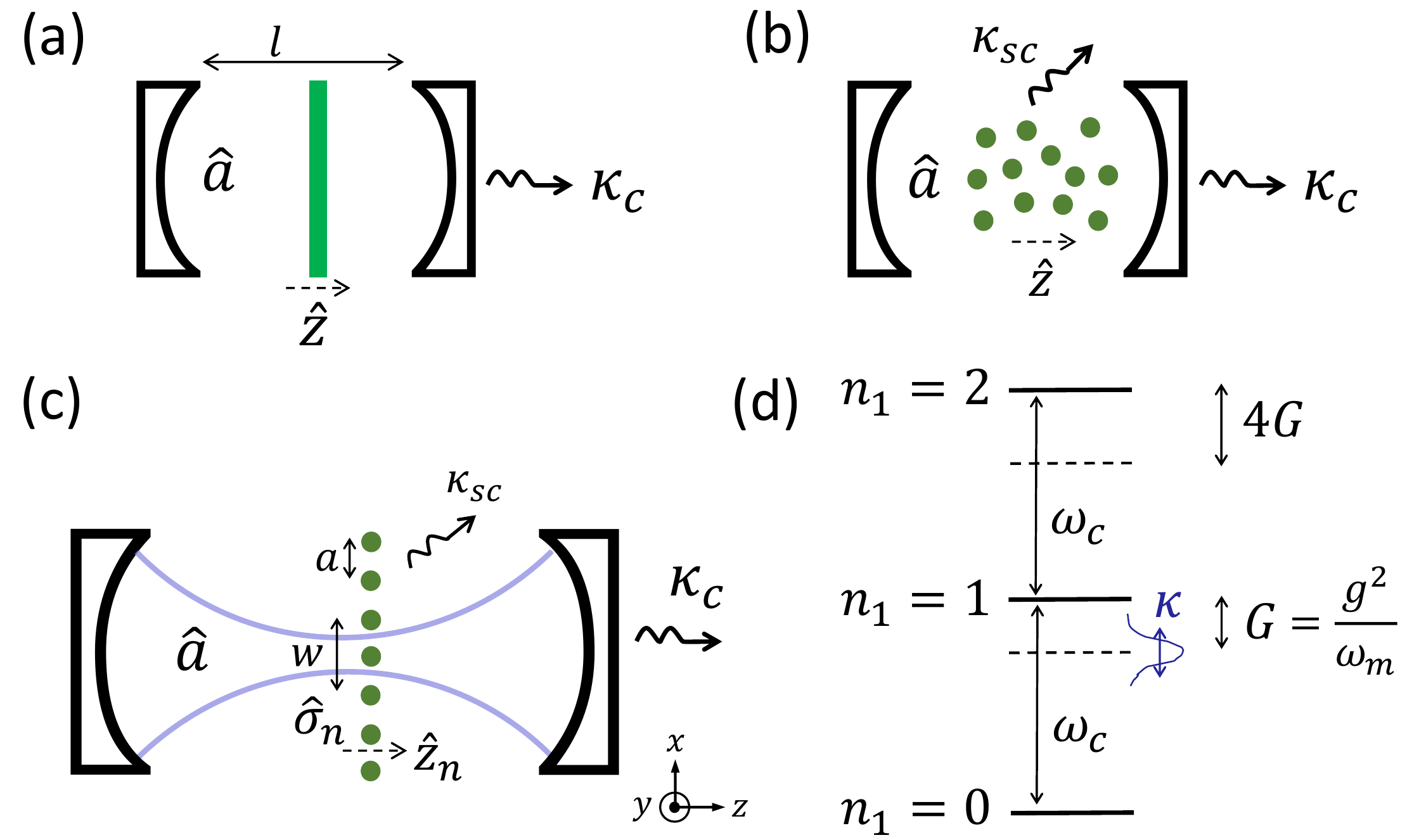}
    \caption{\small{
Cavity optomechanics schemes. (a) Membrane: position $\hat{z}$ affects the resonance frequency of the cavity mode $\hat{a}$ (width $\kappa_c$ due to mirror outcoupling).
(b) Atom-cloud: $\hat{z}$ is the center-of-mass coordinate of the atoms. Total cavity decay, $\kappa=\kappa_c+\kappa_{sc}$, includes transverse photon scattering from atoms $\kappa_{sc}$.
(c) 2D atom-array: $\mathcal{N}$ two-level atoms ($\hat{\sigma}_n$, $n=1,...,\mathcal{N}$) span the $xy$ plane with lattice spacing $a$ ($\mathcal{N}a^2\gg \pi w^2$; $w$ is cavity-mode waist). Motion $\hat{z}_n$ is assumed only along $z$ axis.
(d) Photon energy levels of Hamiltonian (\ref{Hom}) ($\Omega=0$). The levels $n_1=0,1,2,...$ (width $\kappa$) acquire a nonlinear shift $-G n_1^2$.
    }} \label{fig1}
  \end{center}
\end{figure}

Nevertheless, all optomechanical phenomena observed thus far rely on strong optical driving and are limited to Gaussian quantum statistics \cite{AKM}. Few-photon optomechanical nonlinearities, wherein the radiation pressure exerted by a single photon appreciably affects the phase of a subsequent photon \cite{RAB,GIR}, remain unattainable due to inherent limitations of current platforms. For example, the optomechanical coupling of a typical membrane scheme \cite{HAR} (Fig. 1a) is limited by the mechanical susceptibility of the clamped bulk membrane, which is too small to be appreciably pushed by a single photon. Replacing the membrane with a laser-trapped atomic cloud (Fig. 1b) \cite{SK1,SK,ESS,CAM,RES} greatly enhances the optomechanical coupling, however, at the price of increased dissipation caused by transverse scattering from the atoms.

In this Letter, we show how these limitations can be circumvented by using the scheme illustrated in Fig. 1c, wherein the role of the intra-cavity membrane is played by an ordered two-dimensional (2D) array of laser-trapped atoms, such as a 2D optical lattice or tweezer array \cite{OL,QGM,BRW1,LUK,BRW2}. Similar to a disordered atomic cloud, the large mechanical susceptibility of trapped atoms yields very strong optomechanical coupling. The crucial difference, however, is that the spatial order of the array results in a highly directional coupling to light, such that light is scattered almost entirely into the cavity mode and the cavity linewidth remains very small. Therefore, the atom array combines the advantages of both a solid membrane and an atom cloud, namely, directional scattering (low loss) and large optomechanical coupling, respectively. We show that this combination leads to optomechanical parameters that allow to reach the single-photon regime with current technologies. As an illustration, we find the emergence of non-classical correlations such as photon blockade and time-periodic photon correlations mediated by phonons.

\emph{Single-photon nonlinearity via optomechanics.---}
To understand the key idea of this work, we consider the system from Fig. 1a, which can be modeled by the Hamiltonian
\begin{equation}
H=\hbar\omega_c\hat{a}^{\dag}\hat{a}+\hbar g \hat{a}^{\dag}\hat{a}(\hat{b}+\hat{b}^{\dag})+\hbar\omega_m \hat{b}^{\dag}\hat{b}
+\hbar\left(\Omega e^{-i\omega_L t}\hat{a}^{\dag}+\mathrm{h.c.}\right).
\label{Hom}
\end{equation}
Here $\hat{a}$ and $\hat{b}$ are the cavity photon and mechanical phonon bosonic modes with frequencies $\omega_c$ and $\omega_m$, respectively. The corresponding mechanical coordinate is $\hat{z}=x_0(\hat{b}+\hat{b}^{\dag})$, with $x_0$ the zero-point motion. For the case of trapped atoms (Figs. 1b,c), where $x_0=\sqrt{\hbar/(2m\omega_m)}$ with atom mass $m$, we assume motion only along the longitudinal $z$-axis (tight transverse confinement). The optomechanical coupling $g$ is related to the cavity frequency shift per phonon amplitude. The cavity mode has a linewidth $\kappa$ which together with $g$ and $\omega_m$ constitute the three basic optomechanical parameters.

The energy levels of Hamiltonian (\ref{Hom}) (without drive, $\Omega=0$) are given by $E_{n_1 n_2}=\hbar(\omega_c-G n_1)n_1+\hbar\omega_mn_2$, where $G=g^2/\omega_m$ is the nonlinear frequency shift per photon \cite{RAB,GIR}, as depicted in Fig. 1d for photon numbers $n_1=0,1,2$ and a given (non-negative) integer $n_2$.
To observe nonlinearities at the single-photon level, we need a shift $G$ larger than the width $\kappa$. For a photon blockade we further require the sideband-resolved regime \cite{RAB}, leading, respectively, to the two conditions
\begin{subequations}
\noindent\begin{tabularx}{\columnwidth}{@{}XXX@{}}
\begin{equation}
%\frac{g^2}{\omega_m\kappa}\gg 1,
g^2/\omega_m \gg \kappa,
\label{con1}
\end{equation} &
\begin{equation}
%\frac{\omega_m}{\kappa}> 1.
\omega_m>\kappa.
\label{con2}
\end{equation}
\end{tabularx}
\end{subequations}
In the following, we use this pair of requirements as a benchmark for single-photon nonlinearity.

Let us now consider the optomechanical parameters of the different schemes from Fig. 1, summarized in Table I. The optomechanical coupling $g$ is always proportional to the Lamb-Dicke parameter $\eta=qx_0=(\omega_c/c)x_0$  (mechanical trap stiffness). For a suspended bulk membrane (Fig. 1a) we have  $\eta\sim 10^{-7}$ ($x_0\sim 10^{-14}$ m) \cite{AKM,HAR,BAC}, and $g$ is typically much too small to satisfy condition (2a). For the atom cloud from Fig. 1b, where $\hat{z}$ is the center-of-mass coordinate of the laser-trapped atoms, the situation is greatly improved with $\eta \sim 10^{-1}$ ($x_0\sim 10^{-8}$ m) and $g\propto\eta \sqrt{N} \mathrm{Re}\chi$, benefiting also from the atom number $N$ \cite{SK1}. Here $\chi=-\gamma/(2\delta+i\gamma)$ is proportional to the electric susceptibility of the cloud, modeled by two-level atoms with free-space emission rate $\gamma$ and frequency $\omega_a$ ($\delta=\omega_c-\omega_a$ being the cavity-atom detuning). Its imaginary part, related to photon scattering to modes outside of the cavity, gives a substantial contribution to the linewidth, $\kappa_{sc}\propto N \mathrm{Im}\chi$, on top of the intrinsic cavity linewidth $\kappa_c$. This typically leads to the invalidity of condition (2b) [and possibly also (2a)]. Considering the atom-array case (Fig. 1c), we find below that $g$ is similar to that of the atom cloud. However, for small fluctuations in the ordered-array positions ($\eta\ll 1$), interference reduces the outside scattering $\kappa_{sc}$ by a factor $\eta^2$. The resulting favorable scaling $g\propto \eta$ and $\kappa_{sc}\propto \eta^2$, with $\omega_m=E_R/(\hbar\eta^2)$ [$E_R=\hbar^2q^2/(2m)$] and $\eta\sim 0.1$, then allows to satisfy both conditions (2), as illustrated further below.

\begin{table}[ht]
\centering % used for centering table
\begin{tabular}{|c|c|c|c|} % centered columns (5 columns)
\hline %inserts double horizontal lines
%\hline
   & (a) membrane & (b) atom cloud & (c) atom array \\ [0.5ex] % inserts table heading
\hline % inserts single horizontal line
$g$ & \(\displaystyle \sim \eta \frac{c}{l}\)  & \(\displaystyle\eta \frac{c}{l}\frac{\gamma}{\delta}\sqrt{N}\frac{3}{q^2 w^2}\)& \(\displaystyle\eta \frac{c}{l}\frac{\gamma}{\delta-\Delta}\sqrt{N}\frac{3}{q^2 w^2}\)  \\ [2ex]
\hline % inserts single horizontal line
$\kappa$ & \(\displaystyle\kappa_c=\frac{c}{l}\frac{\pi}{F}\) & \(\displaystyle N\frac{c}{l}\frac{\gamma^2}{\delta^2}\frac{3/2}{q^2 w^2}+\kappa_c\)  &  \(\displaystyle\eta^2N\frac{c}{l}\frac{\gamma^2}{(\delta-\Delta)^2}\frac{\alpha/2}{q^2 w^2}+\kappa_c\) \\ [2ex] % [1ex] adds vertical space
\hline %inserts single line
\end{tabular}
\caption{Optomechanical parameters $g$ and $\kappa$ for the different schemes from Fig. 1 (a, b, c). Here $l$ is the cavity length, $F$ its finesse, $w$ the waist of the cavity mode, $\eta$ the Lamb-Dicke parameter, $N$ the number of atoms within the waist ($N=\pi w^2/a^2$ for the array), $q=\omega_c/c$, and $\alpha=21/5$.
The trap equilibrium position $z_0=(M+1/2)\pi/2q$ ($M$ integer) is taken. Compared to the atom cloud, the array exhibits a cooperative shift $\Delta$ and a suppression by $\eta^2$ of outside scattering $\kappa_{sc}=\kappa-\kappa_c$.
}
\label{table} % is used to refer this table in the text
\end{table}

\emph{Atom-array membrane.---}
We now turn to a detailed analysis of the many-particle, multimode atom-array system, showing that it can be mapped to the simple model (\ref{Hom}) with the parameters from Table I(c).  Before we discuss the derivation of this mapping \cite{cQEDf}, we explain its physical origin.

For a square 2D array with lattice spacing $a\lesssim\lambda=2\pi/q$, multiple scattering of light between atoms (dipole-dipole interaction) may become significant and atoms cannot be treated as individual dipoles with the resonant frequency $\omega_a$. Instead, the normal dipole modes for an array in free space become lattice Fourier modes $\mathbf{k}_{\bot}=(k_x,k_y)$ ($k_{x,y}\in\{-\pi/a,\pi/a\}$), with a lowering operator and cooperatively shifted resonance \cite{coop},
\begin{equation}
\hat{\sigma}_{\mathbf{k}_{\bot}}=\frac{1}{\sqrt{\mathcal{N}}}\sum_{n=1}^{\mathcal{N}} e^{-i\mathbf{k}_{\bot}\cdot \mathbf{r}^{\bot}_n}\hat{\sigma}_n
\quad \mathrm{and} \quad
\omega_{\mathbf{k}_{\bot}}=\omega_a+\Delta_{\mathbf{k}_{\bot}},
\label{sig}
\end{equation}
respectively. Here $\hat{\sigma}_n$ is the lowering operator of a two-level atom $n$ at fixed $xy$ lattice point $\mathbf{r}^{\bot}_n$ and $\mathcal{N}$ is the number of array atoms (Fig. 1c). By lattice symmetry, a collective dipole mode $\hat{\sigma}_{\mathbf{k}_{\bot}}$ only couples to propagating photons with the same transverse wavevector $\mathbf{k}_{\bot}$ \cite{note1}. Therefore, to zeroth order in the motion, where all atoms are at their identical longitudinal equilibrium positions, $\hat{z}_n\rightarrow z_0$, and the array is perfectly ordered (Fig. 2a), the transversely confined cavity mode can excite only spatially matched dipole modes, which cannot scatter to outside, non-confined photon modes, resulting in $\kappa_{sc}=0$.
To first order in motion, the mechanical collective coordinate $\hat{z}$, to which the cavity mode couples, inherits the latter's Gaussian profile with waist $w$ (Fig. 2b)
\begin{equation}
\hat{z}=\sum_{n} V_n^0 \hat{z}_n=x_0(\hat{b}+\hat{b}^{\dag}), \quad V_n^0=\frac{2}{\sqrt{\pi}}\frac{a}{w}e^{-2(r_n^{\bot}/w)^2}.
\label{b}
\end{equation}
The resulting coupling between $\hat{b}$ and $\hat{a}$, $g$ from Table I(c), resembles that of the atom-cloud scheme including the additional cooperative shift $\Delta_{\mathbf{k}_{\bot}}\approx \Delta_{\mathbf{0}}\equiv \Delta$ of the paraxial ($\mathbf{k}_{\bot}\rightarrow 0$) collective dipoles excited by the cavity field.
Turning to second order, the differences in longitudinal positions $z_n\neq z_{n'}$ caused by the motion (Fig. 2b) perturb the perfect lattice symmetry, giving rise to coupling between the cavity-confined and the decaying, non-confined dipole modes, and subsequently to outside scattering $\kappa_{sc}$ (reduced by a factor $\eta^2$ compared to the atom-cloud case).
A complementary, conservative second-order optomechanical coupling also exists, captured by the additional Hamiltonian term
\begin{equation}
H_2=\hbar g_2(\hat{b}+\hat{b}^{\dag})^2\hat{a}^{\dag}\hat{a},\quad g_2=\cos(2qz_0)\eta^2\frac{c}{l}\frac{\gamma}{\delta-\Delta}\frac{4}{q^2w^2}.
\label{H2}
\end{equation}
For the equilibrium position $z_0=(M+1/2)\pi/2q$ ($M$ integer) chosen in Table I(c), which maximizes $g$ \cite{cQEDf}, $\cos(2qz_0)=0$ and $H_2$ vanishes (as typical for membrane schemes \cite{HAR}).

\begin{figure}[t]
  \begin{center}
    \includegraphics[width=\columnwidth]{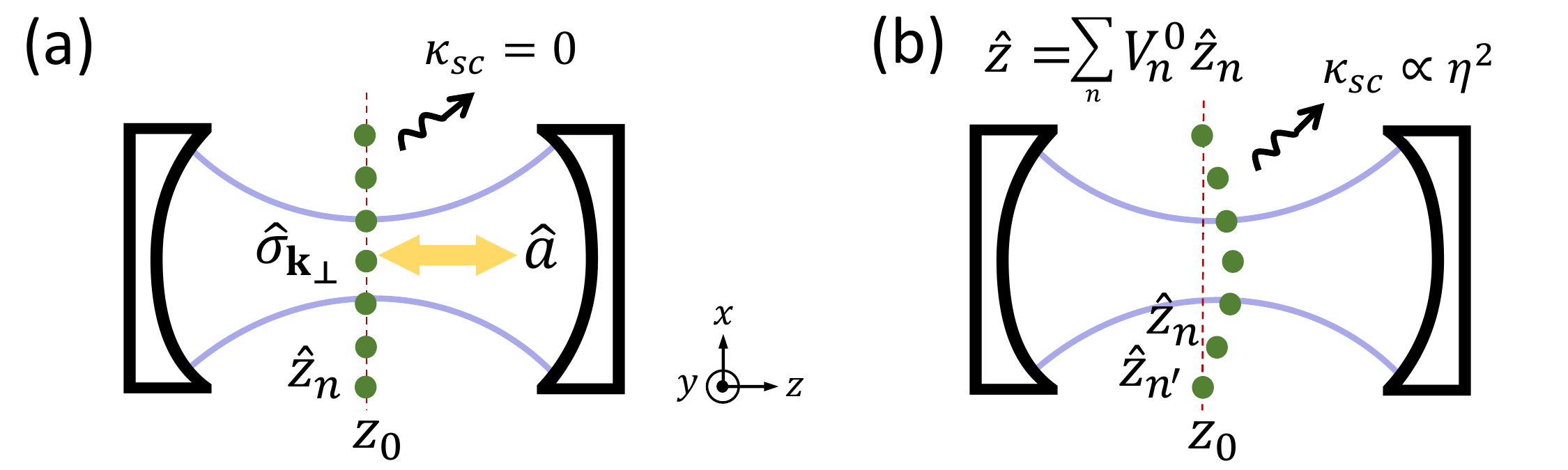}
    \caption{\small{
Atom-array membrane. (a) Without motion the array is perfectly ordered ($z_n=z_0$ $\forall n$, $\mathbf{k}_{\bot}$ conserved by array), and the confined cavity mode $\hat{a}$ interacts only with spatially-matched, confined wavepackets of collective dipoles $\hat{\sigma}_{\mathbf{k}_{\bot}}$, so that scattering to outside, non-confined modes vanishes, $\kappa_{sc}=0$.
(b) To first order in motion, we identify the optomechanically relevant collective coordinate $\hat{z}$, which inherits the Gaussian profile of the cavity mode, $V_n^0$ [Eq. (\ref{b})]. To second order, motion also perturbs the otherwise perfectly-flat array, $z_n \neq z_{n'}$, leading to outside scattering, $\kappa_{sc}\propto \eta^2$.
    }} \label{fig2}
  \end{center}
\end{figure}

Before we discuss the consequences of the optomechanical parameter regime of the atom array, we briefly outline the derivation of the mapping to the simple model (\ref{Hom}) starting from a complex system involving many atoms and photonic modes, following the formalism developed in \cite{cQEDf}. The atoms possess both internal and motional degrees of freedom [$\hat{\sigma}_n$ and $\hat{z}_n=x_0(\hat{b}_n+\hat{b}_n^{\dag})$], while the photon modes, spanning 3D space (mode indices $\{\mu\}$), are divided into transversely confined modes, out of which we consider a single cavity mode $\hat{a}$, and to non-confined modes (``$\mu\in$ nc") $\hat{a}_{\mu}$ with mode functions $u_{\mu}(\mathbf{r})$. The full Hamiltonian is $H=H_A+H_P+H_{AP}$ with $H_A=\hbar\omega_a\sum_n\hat{\sigma}_n^{\dag}\hat{\sigma}_n+\hbar\omega_m \sum_n\hat{b}_n^{\dag}\hat{b}_n$ and
\begin{eqnarray}
H_P&=&\hbar\omega_c\hat{a}^{\dag}\hat{a}+\hbar\left(\Omega e^{-i\omega_L}\hat{a}^{\dag}+\mathrm{h.c.}\right)
+\sum_{\mu\in \mathrm{nc}}\hbar\omega_{\mu}\hat{a}^{\dag}_{\mu}\hat{a}_{\mu},
\nonumber\\
H_{AP}&=&\hbar\sum_n\left[2g_n\sin(q\hat{z}_n)\hat{a}-\sum_{\mu\in \mathrm{nc}}ig_{\mu}(\hat{\mathbf{r}}_n)\hat{a}_{\mu}\right]\sigma^{\dag}_n+\mathrm{h.c.},
\nonumber\\
\label{H}
\end{eqnarray}
where $g_n\propto e^{-(r^{\bot}_n/w)^2}$ and $g_{\mu}(\hat{\mathbf{r}}_n)\propto u_{\mu}(\hat{\mathbf{r}}_n)$ are the dipole couplings to the cavity and non-confined modes, respectively, and $\hat{\mathbf{r}}_n=(\mathbf{r}_n^{\bot},\hat{z}_n)$. The mapping to (\ref{Hom}) is derived using the following steps \cite{cQEDf}.
\\
\emph{(1) Generic treatment of non-confined modes:} Eliminating the modes $\hat{a}_{\mu}$ as a Markovian reservoir, we derive the Heisenberg-Langevin equations for $\hat{\sigma}_n$, $\hat{b}_n$ and $\hat{a}$, including the additional damping of $\hat{a}$ via the mirrors ($\kappa_c$). The resulting equation for $\tilde{\sigma}_n=e^{i\omega_L t}\hat{\sigma}_n$ is (atoms assumed far from saturation),
\begin{eqnarray}
\dot{\tilde{\sigma}}_n=i\delta \tilde{\sigma}_n-i2g_n\sin(q\hat{z}_n)\tilde{a}-\sum_{n'}D(\hat{\mathbf{r}}_n,\hat{\mathbf{r}}_{n'})\tilde{\sigma}_{n'}+\hat{F}_n,
\label{50a}
\end{eqnarray}
where $\hat{F}_n$ is a Langevin noise and the corresponding equations for $\hat{b}_n$ and $\tilde{a}=e^{i\omega_L t}\hat{a}$ can be found in \cite{cQEDf}. Here
$D(\mathbf{r},\mathbf{r}')=-i(3/2)\gamma\lambda \sum_{\mu\in\mathrm{nc}}\frac{u_\mu(\mathbf{r})u^{\ast}_\mu(\mathbf{r}')}{(\omega_\mu/c)^2-q^2}$ is the dipole-dipole interaction kernel mediated by the non-confined modes. A crucial step in our approach is finding an approximation for $D(\mathbf{r},\mathbf{r}')$ which is independent of the specific cavity structure and mode functions $u_\mu(\mathbf{r})$, but that correctly accounts for the directional coupling of the array with light. Specifically, for a collective dipole whose spatial profile matches that of a cavity-confined mode we identify that: (i) the collective dipole does not emit to outside, non-confined modes (Fig. 2a); (ii) its cooperative shift is dominated by near fields and is therefore well approximated by its value in free space $\Delta$. A generic way to satisfy these requirements is to take \cite{cQEDf}
\begin{eqnarray}
D = D^{\mathrm{FS}}-D^{\mathrm{c}}, \quad \mathrm{with} \quad \mathrm{Im}[D]\approx \mathrm{Im}[D^{\mathrm{FS}}],
\label{31}
\end{eqnarray}
respectively, where $D^{\mathrm{FS}}$ is the dipole-dipole kernel in free space and $D^{\mathrm{c}}$ is that mediated by transversely confined modes (e.g. Hermite-Gauss modes).
\\
\emph{(2) Small amplitude motion:} Writing $\hat{z}_n=z_0+\delta\hat{z}_n$ we expand all quantities to second order in $q\delta\hat{z}_n$, finding e.g. $D(\hat{\mathbf{r}}_n,\hat{\mathbf{r}}_{n'})\approx D(\mathbf{r}^{\bot}_n,\mathbf{r}^{\bot}_{n'})+\hat{J}_{nn'}$ with $\hat{J}_{nn'}\propto (\delta\hat{z}_n-\delta\hat{z}_{n'})^2$ \cite{cQEDf}. For Eq. (\ref{50a}) with (\ref{31}) and transforming to collective dipoles (\ref{sig}), we obtain \cite{cQEDf}
\begin{eqnarray}
\dot{\tilde{\sigma}}_{\mathbf{k}_{\bot}}=i(\delta-\Delta_{\mathbf{k}_{\bot}})\tilde{\sigma}_{\mathbf{k}_{\bot}}
-\sum_{\mathbf{k}'_{\bot}}\left[\frac{\gamma_{\mathbf{k}_{\bot}\mathbf{k}'_{\bot}}}{2}+\hat{J}_{\mathbf{k}_{\bot}\mathbf{k}'_{\bot}}\right]\tilde{\sigma}_{\mathbf{k}'_{\bot}}
+\hat{B}_{\mathbf{k}_{\bot}},
\label{36}
\end{eqnarray}
where $\hat{B}_{\mathbf{k}_{\bot}}$ is an expression involving $\tilde{a}$, $\hat{z}_n$ and $\hat{F}_n$ \cite{cQEDf}. The mixing of different $\mathbf{k}_{\bot}$ modes via the zeroth-order-motion kernel $\gamma_{\mathbf{k}_{\bot}\mathbf{k}'_{\bot}}\propto \mathrm{Re}[D(\mathbf{r}^{\bot}_n,\mathbf{r}^{\bot}_{n'})]$ does not lead to outside scattering, as guaranteed by Eq. (\ref{31}). Coupling to non-confined modes is established however via the ``disorder" kernel $\hat{J}_{\mathbf{k}_{\bot}\mathbf{k}'_{\bot}}\propto \hat{J}_{nn'}=\mathcal{O}(\delta\hat{z}_n^2)\propto\eta^2$ (Fig. 2b).
\\
\emph{(3) Large detuning:} Assuming $|\delta-\Delta_{\mathbf{k}_{\bot}}|\gg \gamma_{\mathbf{k}_{\bot}\mathbf{k}'_{\bot}},\dot{\hat{z}}_n/\hat{z}_n,\dot{\tilde{a}}/\tilde{a}$,
we adiabatically eliminate the internal atomic states $\tilde{\sigma}_{\mathbf{k}_{\bot}}$ and obtain coupled equations for $\tilde{a}$ and $\hat{b}_n$.
Moving to orthonormal collective modes $\hat{b}_{\nu}=\sum_n V_n^{\nu}\hat{b}_n$, with the $\nu=0$ mode being that from Eq. (\ref{b}), we find
\begin{eqnarray}
\dot{\hat{b}}_{\nu}=-i\omega_m\hat{b}_{\nu}-ig\hat{a}^{\dag}\hat{a}\delta_{0\nu}-i\sum_{\nu'}\mathcal{M}_{\nu\nu'}\left(\hat{b}_{\nu'}+\hat{b}^{\dag}_{\nu'}\right)\hat{a}^{\dag}\hat{a}.
\nonumber\\
\label{116b}
\end{eqnarray}
Here $g$ is that from Table I(c) including a prefactor $\sin(2qz_0)$, and $\mathcal{M}_{\nu\nu'}\propto \eta^2$ (noting that light-induced friction is negligible in the few-photon regime) \cite{cQEDf}.
\\
\emph{(4) Single mechanical mode:} Eq. (\ref{116b}) reveals that only the mode $\nu=0$ is directly coupled to the cavity, whereas all other modes are coupled to each other and to $\hat{a}$ at order $\eta^2$. Treating this coupling perturbatively, we eliminate the modes $\nu\neq 0$ and obtain coupled equations for $\hat{a}$ and $\hat{b}$, whose conservative part obeys Hamiltonian (\ref{Hom}) [including (\ref{H2})], and with a cavity damping (and associated quantum noise) $\kappa$ from Table I(c) \cite{cQEDf}.

\begin{figure}[t]
  \begin{center}
    \includegraphics[width=\columnwidth]{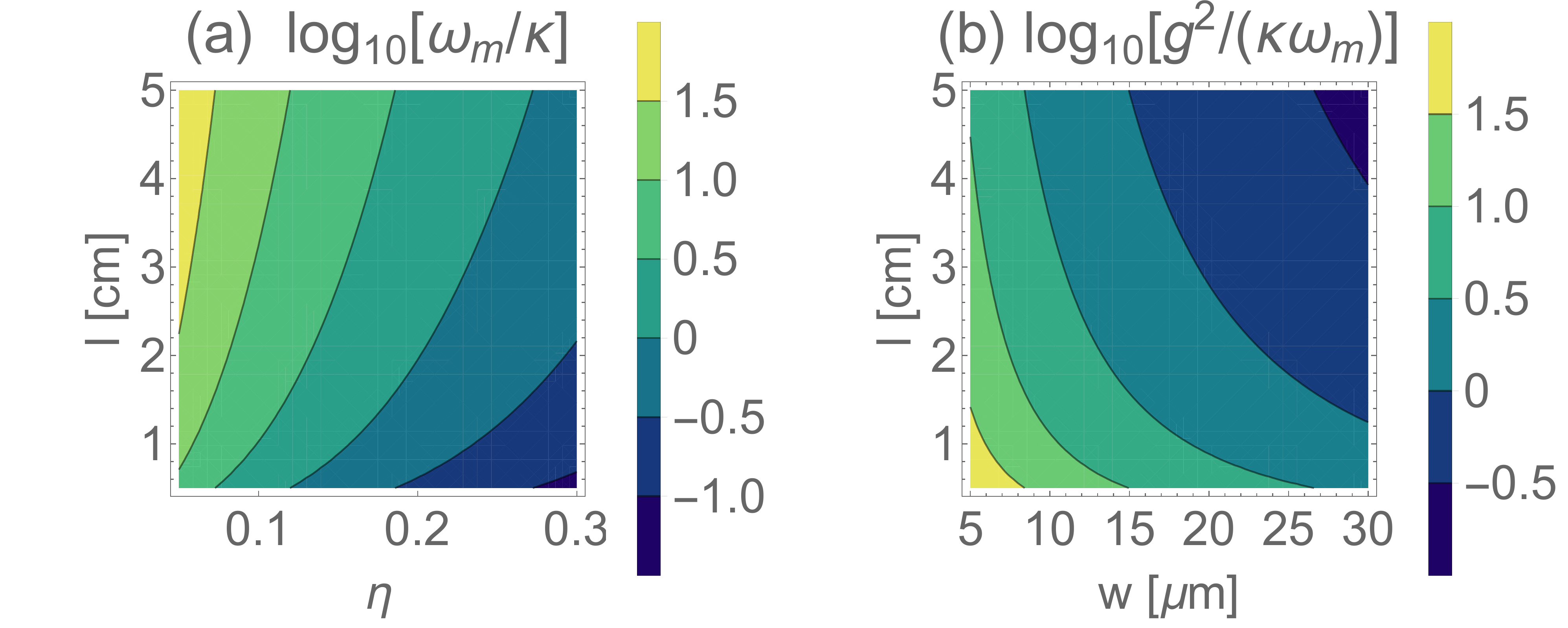}
    \caption{\small{
Realistic optomechanical parameters for an atom array. Using Table I(c) (with $\hbar\omega_m=E_R/\eta^2$), the sideband-resolved ($\omega_m/\kappa$) and blockade ($g^2/\kappa\omega_m$) parameters from conditions (2) are plotted as a function of realistic physical-system parameters. The single-photon regime (values above $0$ in both plots) can be reached for a variety of cavity lengths $l$, mode waists $w$ and Lamb-Dicke parameters $\eta$. Here $F=150000$, $a/\lambda=0.6$, $(\delta-\Delta)/\gamma=(150/4\pi)(\lambda/a)^2$, $\lambda\approx 800$ nm, and $\gamma=1620E_R/\hbar=2\pi\times6$ MHz. In (b) $\eta=0.15$ (i.e. $\omega_m/2\pi=165$ kHz).
    }} \label{fig3}
  \end{center}
\end{figure}

\emph{Ultrastrong quantum optomechanics.---}
With the mapping to (\ref{Hom}) and the expressions for $g$, $\kappa$ and $\omega_m=E_R/(\hbar\eta^2)$ in hand [Table I(c)], we present in Fig. 3 the combined parameters from conditions (2), as a function of physical parameters of realistic cavity and atom-array systems. In all plots we use finesse $F=150000$ \cite{SK,HEM}, lattice spacing $a/\lambda=0.6$, detuning $(\delta-\Delta)/\gamma=(150/4\pi)(\lambda/a)^2$, $\lambda\approx 800$ nm, and $\gamma=1620E_R/\hbar=2\pi\times6$ MHz (the latter for $^{87}$Rb atoms). Beginning with the resolved sideband condition (\ref{con2}), we observe in Fig. 3a that it is achievable with a centimeter-scale cavity length $l$ and a variety of trapping strengths $\eta$. In Fig. 3b we plot the blockade condition (\ref{con1}) for $\eta=0.15$ (corresponding to $\omega_m/2\pi=165$ kHz) as a function of the length $l$ and mode waist $w$ and find that it is satisfied for waists up to at least $30$ $\mu$m.

\begin{figure}[t]
  \begin{center}
    \includegraphics[width=\columnwidth]{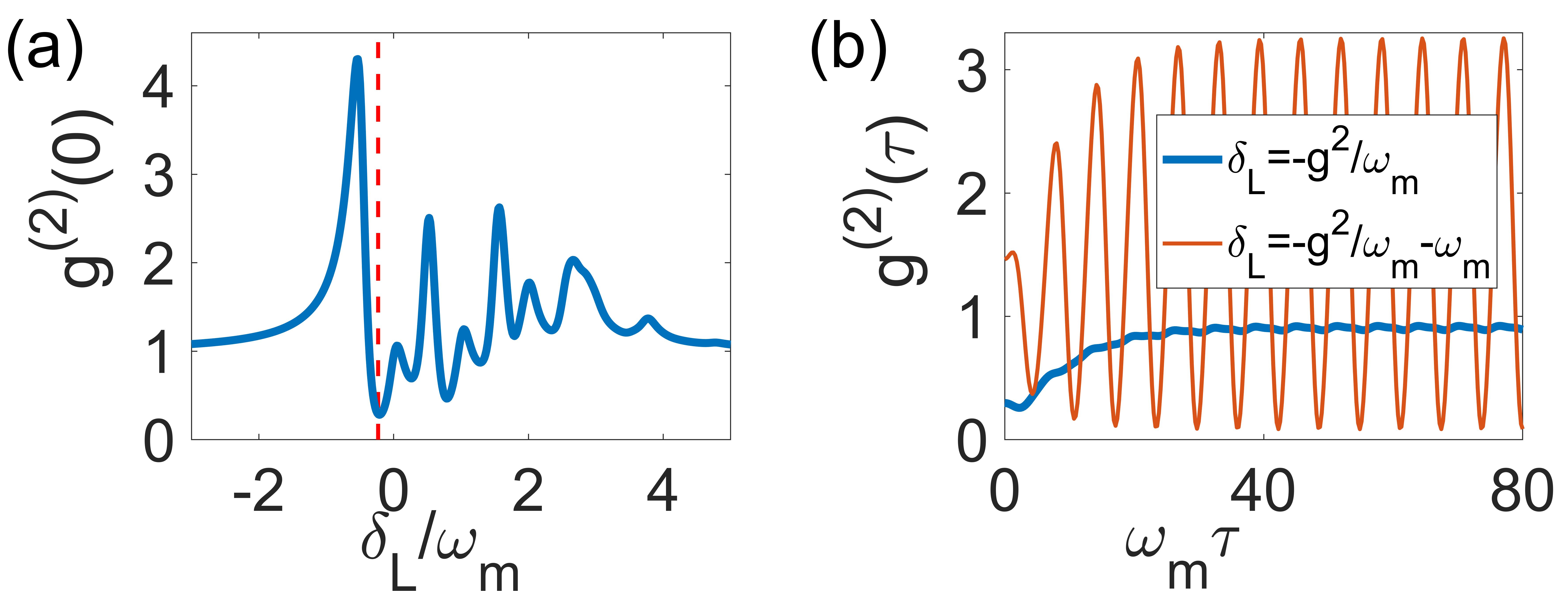}
    \caption{\small{
Non-classical photon statistics in the ultrastrong regime ($g/\omega_m=0.49$ and $\kappa/\omega_m=0.275$; see text for corresponding physical-system parameters).
(a) $g^{(2)}(0)$ as a function of laser-cavity detuning $\delta_L=\omega_L-\omega_c$. Strong antibunching (photon blockade) is observed at several detunings, with the optimum around $\delta_L=-g^2/\omega_m$ (vertical dashed line). Additional resonances result from multiphonon excitations.
(b) $g^{(2)}(\tau)$ for $\delta_L=-g^2/\omega_m$ (blue) and $\delta_L=-g^2/\omega_m-\omega_m$ (red). In the former, antibunched correlations at $\tau=0$ relax to a value below $1$ due to phonon excitation and memory. In the latter, non-classical correlations, $g^{(2)}(\tau)>g^{(2)}(0)>1$, are exhibited at times $\tau$ with the phonon period $2\pi/\omega_m$.
    }} \label{fig4}
  \end{center}
\end{figure}

As an illustration of observable single-photon effects, we choose a specific set of physical parameters $\eta=0.15$, $w=15$ $\mu$m, and $l=3.2$ cm, yielding the model parameters $g/\omega_m=0.49$ and $\kappa/\omega_m=0.275$.
Using a quantum trajectories semi-analytic approach \cite{SI}, we calculate the dynamics of Hamiltonian (\ref{Hom}) with cavity damping $\kappa$ and weak driving $\Omega\ll \kappa$, finding the second-order coherence of the cavity field
$g^{(2)}(\tau)=\frac{\langle \hat{a}^{\dag}(t)\hat{a}^{\dag}(t+\tau)\hat{a}(t+\tau)\hat{a}(t)\rangle}{\langle\hat{a}^{\dag}\hat{a}\rangle^2(t)}$
in steady-state, for an initial vacuum state of the cavity and motion. In Fig. 4a we plot $g^{(2)}(0)$ as a function of the laser-cavity detuning, $\delta_L=\omega_L-\omega_c$, finding antibunching, $g^{(2)}(0)<1$, and hence photon blockade. For the optimal detuning $\delta_L=-g^2/\omega_m$ \cite{RAB}, we also plot the time-delayed correlation $g^{(2)}(\tau)$ and observe its relaxation at a time scale $1/\kappa$ from antibunching towards small steady oscillations at mechanical frequency $\omega_m$ around a value lower than $1$. This behavior signifies the excitation of the collective phonon mode $\hat{b}$ of the array and the resulting memory of photon correlations it mediates. This memory effect is absent in typical photon nonlinearities via two-level systems, and can lead to non-classical features beyond photon-blockade. For example, choosing the detuning $\delta=-g^2/\omega_m-\omega_m$, where antibunching does not occur $g^{(2)}(0)>1$, we observe violation of classicality, $g^{(2)}(\tau)>g^{(2)}(0)>1$ \cite{g2}, at delay-times $\tau$ with a period $2\pi/\omega_m$.

We note that these quantum effects can be observed by modifying established atom-cloud systems if the atoms are trapped in a lattice instead of in a disordered cloud. For example, results similar to Fig. 4 are achievable with the system reported in Ref. \cite{HEM} (wherein $F=340000$, $w=30$ $\mu$m, and $l=5$ cm) for atoms loaded to a lattice with $a/\lambda=0.68$ ($532$nm$/780$nm) and $\eta=0.2$ ($\omega_m/2\pi=92$ kHz) \cite{cQEDf}.

\emph{Discussion.---}
By combining the strong mechanical susceptibility with the cooperative directional scattering of an atom array, this work paves the way for the exploration of few-photon regimes of optomechanics. Considering the utility of single-photon nonlinearities in quantum information \cite{CHAr,KD,MPS,REM,LIN,GER,KIM1,DAY,LUK1,FRI,FIR}, it could be interesting to explore how the memory associated with the phonon-induced nonlinearity discussed above may be exploited to generate complex many-body entangled states \cite{MPS,HAN}. Moreover, it is natural to extend the study of the single-photon regime to the case where the second-order coupling (\ref{H2}) becomes dominant.

Qualitatively new opportunities are opened by considering multiple internal atomic states. In contrast to a solid membrane, characterized by a fixed dielectric response, the atom-array membrane is made of a quantum-coherent ``dielectric" whose internal, many-body state is affected by the optomechanical nonlinearity. For example, the non-classical statistics revealed for the cavity field could potentially be mapped to internal atomic states. This extends the study of optomechanics to consider the backaction, not only on light, but also on the internal quantum state of the membrane. Furthermore, this backaction may be enhanced by operating our proposed scheme at cooperative resonance ($\delta=\Delta$), where a variety of collective dipolar effects are predicted even without a cavity \cite{coop,ADM,janos,ADM2,ANA,CHA2,ABA,ZOL1,ZOL2,BLOe}, including novel multimode optomechanics \cite{om,AAMO}.

\begin{acknowledgments}
We acknowledge fruitful discussions with Peter Rabl, and financial support from the NSF, the MIT-Harvard Center for Ultracold Atoms, the Vannevar Bush Faculty Fellowship, and a research grant from the
Center for New Scientists at the Weizmann Institute of Science.
\end{acknowledgments}


\begin{thebibliography}{}
%cavity optomech general
\bibitem{AKM} M. Aspelmeyer, T. J. Kippenberg and F. Marquardt, Rev. Mod. Phys. \textbf{86}, 1391 (2014).
\bibitem{MEY} P. Meystre, Ann. Phys. \textbf{525}, 215 (2013).
\bibitem{DOR} A. Dorsel, J. D. McCullen, P. Meystre, E. Vignes, and H. Walther, Phys. Rev. Lett. \textbf{51}, 1550 (1983).
\bibitem{HAR} J. D. Thompson, B. M. Zwickl, A. M. Jayich, F. Marquardt, S. M. Girvin and J. G. E. Harris, Nature \textbf{452}, 72 (2008).
%Q technologies with optomech
\bibitem{CAV} C. M. Caves, Phys. Rev. D \textbf{23}, 1693 (1981).
\bibitem{LIGO} J. Aasi \emph{et al.}, Nat. Photon. \textbf{7}, 613 (2013).
\bibitem{LEH} R. W. Andrews, R. W. Peterson, T. P. Purdy, K. Cicak, R. W. Simmonds, C. A. Regal, and  K. W. Lehnert, Nat. Phys. \textbf{10}, 321 (2014).
%cavity optomech cooling
\bibitem{MAR} F. Marquardt, J. P. Chen, A. A. Clerk, and S. M. Girvin, Phys. Rev. Lett. \textbf{99}, 093902 (2007).
\bibitem{WIL} I. Wilson-Rae, N. Nooshi, W. Zwerger, and T. J. Kippenberg, Phys. Rev. Lett. \textbf{99}, 093901 (2007).
\bibitem{SCH} A. Schliesser, O. Arcizet, R. Rivi\`{e}re, G. Anetsberger and T. J. Kippenberg, Nat. Phys. \textbf{5}, 509 (2009).
\bibitem{CHAN} J. Chan, T. P. Mayer Alegre, A. H. Safavi-Naeini, J.T. Hill, A. Krause, S. Gr\"{o}blacher, M. Aspelmeyer and O. Painter, Nature \textbf{478}, 89 (2011).
\bibitem{TEU} J. D. Teufel, T. Donner, D. Li, J. W. Harlow, M. S. Allman, K. Cicak, A. J. Sirois, J. D. Whittaker, K. W. Lehnert and R. W. Simmonds, Nature \textbf{475}, 359 (2011).
%nonclassical states
\bibitem{HAM} K. Hammerer, C. Genes, D. Vitali, P. Tombesi, G. Milburn, C. Simon and D. Bouwmeester, arXiv:1211.2594 (2012).
\bibitem{SK3} D. W. C. Brooks, T. Botter, S. Schreppler, T. P. Purdy, N. Brahms and D. M. Stamper-Kurn, Nature \textbf{488}, 476 (2012).
\bibitem{REG} T. P. Purdy, P.-L. Yu, R. W. Peterson, N. S. Kampel, and C. A. Regal, Phys. Rev. X \textbf{3}, 031012 (2013).
\bibitem{SAF} A. H. Safavi-Naeini, S. Gr\"{o}blacher, J. T. Hill, J. Chan, M. Aspelmeyer and O. Painter, Nature \textbf{500}, 185  (2013).
\bibitem{SCH2} W. H. P. Nielsen, Y. Tsaturyan, C. B. M{\o}ller, E. S. Polzik and Albert Schliesser, Proc. Natl. Acad. Sci. USA \textbf{114}, 62 (2016).
\bibitem{KIP} V. Sudhir, D. J. Wilson, R. Schilling, H. Schütz, S. A. Fedorov, A. H. Ghadimi, A. Nunnenkamp and T. J. Kippenberg, Phys. Rev. X \textbf{7}, 011001 (2017).
\bibitem{SIL1} C. F. Ockeloen-Korppi, E. Damsk\"{a}gg, J.-M. Pirkkalainen, T. T. Heikkil\"{a}, F. Massel and M. A. Sillanp\"{a}\"{a}, Phys. Rev. Lett. \textbf{118}, 103601 (2017).
\bibitem{SIL2} C. F. Ockeloen-Korppi, E. Damsk\"{a}gg, G. S. Paraoanu, F. Massel and M. A. Sillanp\"{a}\"{a}, Phys. Rev. Lett. \textbf{121}, 243601 (2018).
%single-photon optomech
\bibitem{RAB} P. Rabl, Phys. Rev. Lett. \textbf{107}, 063601 (2011).
\bibitem{GIR} A. Nunnenkamp, K. B{\o}rkje and S. M. Girvin, Phys. Rev. Lett. \textbf{107}, 063602 (2011).
%cavity optomech with atoms
\bibitem{SK1} D. Stamper-Kurn, arXiv:1204.4351 (2012).
\bibitem{SK} S. Gupta, K. L. Moore, K. W. Murch and D. Stamper-Kurn, Phys. Rev. Lett. \textbf{99}, 213601 (2007).
\bibitem{ESS} F. Brennecke, S. Ritter, T. Donner and T. Esslinger, Science \textbf{322}, 235 (2008).
\bibitem{CAM} S. Camerer, M. Korppi, A. J\"{o}ckel, D. Hunger, T. W. H\"{a}nsch and P. Treutlein, Phys. Rev. Lett. \textbf{107}, 223001 (2011).
\bibitem{RES} J. Restrepo, C. Ciuti and I. Favero, Phys. Rev. Lett. \textbf{112}, 013601 (2014).

\bibitem{OL} I. Bloch, J. Dalibard and S. Nascimb\`{e}ne, Nat. Phys. \textbf{8}, 267 (2012).
\bibitem{QGM} S. Kuhr, Natl Sci. Rev. \textbf{3}, 170 (2016).
\bibitem{BRW1} D. Barredo, S. de L\'{e}s\'{e}leuc, V. Lienhard, T. Lahaye, A. Browaeys, Science \textbf{354}, 1021 (2016).
\bibitem{LUK} M. Endres, H. Bernien, A. Keesling, H. Levine, E. R. Anschuetz, A. Krajenbrink, C. Senko, V. Vuletic, M. Greiner, M. D. Lukin, Science \textbf{354}, 1024 (2016).
\bibitem{BRW2} D. Barredo, V. Lienhard, S. de L\'{e}s\'{e}leuc, T. Lahaye, and A. Browaeys, Nature \textbf{561}, 79 (2018).

%\bibitem{note} Note however that some signatures of the single-photon regime exist also for $\omega_m<\kappa$ \cite{GIR}.
\bibitem{BAC} P. Weber, J. G\"{u}ttinger, A. Noury, J. Vergara-Cruz and A. Bachtold, Nat. Comm. \textbf{7}, 12496 (2016).   %membrane optomechanics example

\bibitem{cQEDf} E. Shahmoon, D. Wild, M. Lukin and S. Yelin, "Theory of cavity QED with 2D atomic arrays" (jointly submitted with the current work).
\bibitem{coop} E. Shahmoon, D. Wild, M. Lukin and S. Yelin, Phys. Rev. Lett. \textbf{118}, 113601 (2017).
\bibitem{note1} Higher diffraction (Bragg) orders are suppressed for sufficiently small $a\lesssim \lambda$ \cite{coop}.
\bibitem{HEM} M. Wolke, J. Klinner, H. Ke{\ss}ler, and A. Hemmerich, Science \textbf{337}, 75 (2012).
\bibitem{SI} see Supplemental Material, which includes Ref. \cite{QN}.
\bibitem{QN} C. W. Gardiner and P. Zoller, \emph{Quantum Noise}, 3rd Edition (Springer-Verlag, Berlin Heidelberg, Berlin, 2004).
\bibitem{g2} P. K\'{o}m\'{a}r, S. D. Bennett, K. Stannige, S. J. M. Habraken, P. Rabl, P. Zoller, and M. D. Lukin, Phys. Rev. A \textbf{87}, 013839 (2013).

%single-photon nonlinearity in QI
\bibitem{CHAr} D. E. Chang, V. Vuleti\'{c}, and M. D. Lukin, Nat. Photon. \textbf{8}, 685 (2014).
\bibitem{KD} L. M. Duan and H. J. Kimble, Phys. Rev. Lett. \textbf{92}, 127902 (2004).
\bibitem{MPS} C. Sch\"{o}n, E. Solano, F. Verstraete, J. I. Cirac, and M. M. Wolf, Phys. Rev. Lett. \textbf{95}, 110503 (2005).
\bibitem{REM} B. Hacker, S. Welte, G. Rempe, and S. Ritter, Nature \textbf{536}, 193 (2016).
\bibitem{LIN} N. H. Lindner and T. Rudolph, Phys. Rev. Lett. \textbf{103}, 113602 (2009).
\bibitem{GER} I. Schwartz, D. Cogan, E. R. Schmidgall, Y. Don, L. Gantz, O. Kenneth, N. H. Lindner, and D. Gershoni, Science \textbf{354}, 434 (2016).
\bibitem{KIM1} T. Aoki, A. S. Parkins, D. J. Alton, C. A. Regal, B. Dayan, E. Ostby, K. J. Vahala, and H. J. Kimble, Phys. Rev. Lett. \textbf{102}, 083601 (2009).
\bibitem{DAY} I. Shomroni, S. Rosenblum, Y. Lovsky, O. Bechler, G. Guendelman, and B. Dayan, Science \textbf{345}, 903 (2014).
\bibitem{LUK1} T. G. Tiecke, J. D. Thompson, N. P. de Leon, L. R. Liu, V. Vuleti\'{c}, and M. D. Lukin, Nature \textbf{508}, 241 (2014).
\bibitem{FRI} I. Friedler, D. Petrosyan, M. Fleischhauer, and G. Kurizki, Phys. Rev. A \textbf{72}, 043803 (2005).
\bibitem{FIR} O. Firstenberg, T. Peyronel, Q.-Y. Liang, A. V. Gorshkov, M. D. Lukin, and V. Vuleti\'{c}, N. \textbf{502}, 71 (2013).
\bibitem{HAN} H. Pichler, S. Choi, P. Zoller, and M. D. Lukin, Proc. Natl. Acad. Sci. USA \textbf{114}, 11362 (2017).

%\bibitem{force} references on force sensing with ions/atoms or squeezed mech states?

\bibitem{ADM} R. J. Bettles, S. A. Gardiner and C. S. Adams, Phys. Rev. Lett. \textbf{116}, 103602 (2016).
\bibitem{janos} J. Perczel, J. Borregaard, D. E. Chang, H. Pichler, S. F. Yelin, P. Zoller and M. D. Lukin, Phys. Rev. Lett. \textbf{119}, 023603 (2017).
\bibitem{ADM2} R. J. Bettles, J. Min\'{a}\v{r}, C. S. Adams, I. Lesanovsky and B. Olmos, Phys. Rev. A \textbf{96}, 041603(R) (2017).
\bibitem{ANA} A. Asenjo-Garcia, M. Moreno-Cardoner, A. Albrecht, H. J. Kimble, and D. E. Chang, Phys. Rev. X \textbf{7}, 031024 (2017).
\bibitem{CHA2} M. T. Manzoni, M. Moreno-Cardoner and A. Asenjo-Garcia, J. V. Porto, A. V. Gorshkov, and D. E. Chang, N. J. Phys. \textbf{20}, 083048 (2018).
\bibitem{ABA} V. Mkhitaryan, L. Meng, A. Marini, F. J. Garcia de Abajo, Phys. Rev. Lett. \textbf{121}, 163602 (2018)
\bibitem{ZOL1} A. Grankin, P. O. Guimond, D. V. Vasilyev, B. Vermersch, and P. Zoller, Phys. Rev. A \textbf{98}, 043825 (2018).
\bibitem{ZOL2} P.-O. Guimond, A. Grankin, D. V. Vasilyev, B. Vermersch, and P. Zoller, Phys. Rev. Lett. \textbf{122}, 093601 (2019).
\bibitem{BLOe} J. Rui, D. Wei, A. Rubio-Abadal, S. Hollerith, J. Zeiher, D. M. Stamper-Kurn, C. Gross and I. Bloch, arXiv:2001.00795.
\bibitem{om} E. Shahmoon, M. Lukin and S. Yelin, arXiv:1810.01052.
\bibitem{AAMO} E. Shahmoon, M. D. Lukin, and S. F. Yelin, \emph{Advances In Atomic, Molecular, and Optical Physics}, 68, 1 (Elsevier, New York, 2019).
\end{thebibliography}
\end{document}